\documentclass[twocolumn,english,prb,aps,preprintnumbers,amsmath,amssymb,array]{revtex4}
\usepackage[T1]{fontenc}
\usepackage[latin9]{inputenc}
\usepackage{graphicx}

\makeatother
\usepackage{babel}

\begin{document}

\title{Analytic continuation average spectrum method for quantum liquids}

\author{David R. Reichman}
\address{Department of Chemistry, Columbia University, 3000 Broadway, New
York, New York 10027}

\author{Eran Rabani}
\address{School of Chemistry, The Sackler Faculty of Exact Sciences, Tel Aviv
University, Tel Aviv 69978, Israel}

\date{\today}

\begin{abstract}
We revisit the problem of determining the real-frequency density
response in quantum fluids via analytical continuation of
imaginary-time quantum Monte Carlo data.  We demonstrate that the
average spectrum method (ASM) is capable of revealing resolved modes
in the dynamic structure factor of both {\em ortho}-deuterium and
liquid {\em para}-hydrogen, in agreement with experiments and quantum
mode-coupling theories, while the maximum entropy approach yields only
a smooth unimodal spectrum.  Outstanding issues are discussed.  Our
work provides the first application of the ASM method in realistic
off-lattice systems.
\end{abstract}

\maketitle
\newpage{}

\section{Introduction}
\label{sec:intro}
The computation of real-time correlation functions in condensed phase
quantum systems is a challenging problem for which there is no
universally applicable approach. This fact has necessitated the
development of a variety of powerful methods ranging from exact
numerical approaches to approximate
theories.\cite{Rossky94,Voth96review,Makri99,Egorov99f,Tully2000,Hammes-Schiffer2001,Krilov01c,Miller2001,Manolopoulos04,Thoss2004,Rabanireview05,Kapral2006,Kapral2006a,Prezhdo2007,NRG,DMFT,DMRG}
The choice of method will in general depend on the type,
dimensionality, and size of the system of interest as well as the
external thermodynamic conditions under consideration.

The problems that plague exact real-time approaches are generally
absent in imaginary time.\cite{Berne86} The dynamical sign problem,
which renders most direct real time Monte Carlo simulations
infeasible, is avoided in imaginary time. In principle, this means
that a well defined Wick rotation may be used to recover real time
information and excitation spectra.\cite{Mermin} If the precise
analytical form of the correlation function is known in imaginary
time, rotation in the complex time plane is straight forward. However
in most cases of interest the imaginary time correlation function must
be determined numerically via quantum Monte Carlo. In practice,
analytic continuation in this case is numerically unstable and highly
sensitive to statistical error.\cite{Gubernatis96}

The most commonly used technique for the numerical analytical
continuation of imaginary time quantum Monte Carlo data is the maximum
entropy (MaxEnt)
method.\cite{Gubernatis90,Gubernatis91,Gubernatis96,Berne94a,Ceperley96,Berne96a,Krilov99,Rabani00,Krilov01a,Krilov01b,Rabani02c,Rabani03,Rabani05,Manolopoulos07,Miller08,Voth2008}
In MaxEnt, the optimal fit of the data is defined in a Bayesian manner
as the most likely fit that emerges from the competition between the
$\chi^{2}$ goodness-of-fit and an entropic prior $S$. MaxEnt generally
requires an additional means to determine the prefactor $\alpha$
(``temperature'') of the entropy term. Once determined, MaxEnt
provides a statistically rigorous and unique fit for the spectral
function. The strength of the MaxEnt method is that it generally
provides good estimates of spectral area in the correct region of
frequency space. However numerous studies have shown that, in general,
MaxEnt solutions tend to be too broad and smooth and often lack
clearly defined peaks.\cite{Gallicchio98,Rabani04} Other approaches
have been devised to overcome this ``smoothness'' problem but these
approaches have not met with much general success.

While MaxEnt provides a Bayesian estimate of the most likely fit, a
different approach may be devised based on the notion of averaging
over a sequence of possible solutions. Such approaches are called
stochastic analytic continuation methods or average spectrum methods
(ASM).\cite{Sandvik98,Assaad08,Syljuasen08} One justification for such
a procedure is the fact that MaxEnt solutions for different values of
$\alpha$ may have spurious features that vanish or are diminished when
averaged. This type of approach was independently suggested for the
continuation of imaginary time quantum Monte Carlo data by White and
Sandvik,\cite{White,Sandvik98} while similar ideas have been used in
other fields.\cite{Mosegaard,Drummond} While these methods have been
put forward on rather heuristic grounds, the results generated in
several studies give support to the notion that features not detected
by MaxEnt may be faithfully revealed. Beach has provided a more secure
theoretical foundation for such an approach by constructing a version
of an ASM for which MaxEnt is the mean field limit.\cite{Beach} Thus,
one may view the ASM as including fluctuations not incorporated in the
MaxEnt method.

In this work, we will use a variant of the ASM approach outlined by
Slyjuasen~\cite{Syljuasen08} to calculate the dynamic structure factor
of both liquid {\em ortho}-deuterium and liquid {\em
para}-hydrogen. Such cases are useful because extensive experimental
results exist in these systems that show features that are not
captured by MaxEnt. While the ASM approach used here is more
computationally demanding than MaxEnt, we find that this approach is
capable of revealing such subtle features.  Not surprisingly, the
approach fares better for higher quality of the Monte Carlo data.  Our
work provides the first application of the ASM for analytical
continuation of quantum Monte Carlo data for off-lattice systems with
continuous potentials.

The outline of the paper is as follows: In Sec.~\ref{sec:asm} we
outline the average spectrum method. Sec.~\ref{sec:res} describes the
model of liquid {\em ortho}-deuterium and liquid {\em para}-hydrogen,
provides the technical Monte Carlo and inversion details, and presents
the results for the dynamics structure factor of both liquids.  In
Sec.~\ref{sec:conclusions} we conclude.

\section{Analytic continuation average spectrum method}
\label{sec:asm}
The analytic continuation of the intermediate scattering
function~\cite{Ceperley96,Rabani04,Manolopoulos07,Miller08a,Miller08}
is based on the Fourier relation between the dynamic structure factor
$S(q,\omega)$ and the intermediate scattering function $F(q,t)$:
\begin{equation}
F(q,t)=\frac{1}{2\pi}\int_{-\infty}^{\infty}d\omega\mbox{e}^{-i\omega
t}S(q,\omega).
\label{eq:fqt1}
\end{equation}
By performing the replacement $t\rightarrow-i\tau$, and using the
detailed balance relation $S(q,-\omega)=e^{-\beta\omega}S(q,\omega)$
we obtain
\begin{equation}
\tilde{F}(q,\tau)=\frac{1}{2\pi}\int_{0}^{\infty}d\omega\biggl[e^{-\omega\tau}+e^{(\tau-\beta)\omega}\biggr]S(q,\omega),\label{eq:fqtau1}\end{equation}
where $t,\tau\ge0$, and \begin{equation}
\tilde{F}(q,\tau)=\frac{1}{Z}\frac{1}{N}\mbox{Tr}\left(e^{-\beta
,H}e^{\tau H}\hat{\rho}_{{\bf q}}^{\dagger}e^{-\tau H}\hat{\rho}_{{\bf
q}}\right).
\label{eq:fqtau}
\end{equation}
Here, $Z$ is the partition function, the quantum collective density
operator is given by $\hat{\rho}_{{\bf
q}}=\sum_{\alpha=1}^{N}\mbox{e}^{i{\bf q}\cdot{\bf
\hat{r}}_{\alpha}}$, $N$ is the number of particles, ${\bf q}$ the
wave number, and $\hat{r}_{\alpha}$ is the position operator of
particle $\alpha$.

The reason for introducing the imaginary time intermediate scattering
function, $\tilde{F}(q,\tau)$, is that, unlike its real time
counterpart, it is straightforward to obtain it using an appropriate
path-integral Monte Carlo (PIMC) simulation
technique.\cite{Berne86ra,Berne86rb} However, in order to obtain the
dynamic structure factor and the real time intermediate scattering
function one has to invert the integral in Eq.~(\ref{eq:fqtau1}). Due
to the singular nature of the integration kernel the inversion of
Eq.~(\ref{eq:fqtau1}) is an ill-posed problem.  As a consequence, a
direct approach to the inversion would lead to an uncontrollable
amplification of the statistical noise in the data for
$\tilde{F}(q,\tau)$, resulting in an infinite number of solutions that
satisfy Eq.~(\ref{eq:fqtau1}).

In the present study we apply the recently developed averaged spectrum
method,\cite{Syljuasen08} related to earlier stochastic continuation
approaches,\cite{Sandvik98,Beach,Assaad08} to invert the integral in
Eq.~(\ref{eq:fqtau1}) to get $S(q,\omega)$. Following the notation of
other analytic continuation methods, we will refer to
$\tilde{F}(q,\tau)$ as the data (input data generated from the PIMC
approach), $K(\tau,\omega)=e^{-\omega\tau}+e^{(\tau-\beta)\omega}$ as
the singular kernel, and $S(q,\omega)$ as the solution. Furthermore,
in what follows, we will assume that both the imaginary time axis and
the frequency axis are discretized such that $\tau=j_{\tau}\delta\tau$
($j_{\tau}=1,\cdots N_{\tau}$) and $\omega=j_{\omega}\delta\omega$
($j_{\omega}=1,\cdots N_{\omega}$). Hence, Eq.~(\ref{eq:fqtau1}) in
its discrete form is given by:
\begin{equation}
\tilde{F}_{j_{\tau}}(q)=\sum_{j_{\omega}=1}^{N_{\omega}}\delta\omega
K_{j_{\tau},j_{\omega}}S_{j_{\omega}}(q),
\label{eq:fqtau-d}
\end{equation}
where for future reference, the vectors $\tilde{{\bf F}}(q)$ and ${\bf
S}(q)$ describe the data and the solution in discrete space, with
values given by $\tilde{F}_{j_{\tau}}(q)$ and $S_{j_{\omega}}(q)$,
respectively.

The basic idea behind the ASM is to pick the final solution for ${\bf
S}(q)$ as the average spectral function obtained by averaging over a
posterior distribution:
\begin{equation} \bar{{\bf S}}(q)=\frac{\int
d|{\bf S}(q)|{\bf S}(q)P({\bf S}(q)|\tilde{{\bf F}}(q))}{\int d|{\bf
S}(q)|P({\bf S}(q)|\tilde{{\bf F}}(q))},
  \label{eq:sqwav}
\end{equation}
where ${\bf S}(q)$ is a solution to Eq.~(\ref{eq:fqtau-d}) for a given
input $\tilde{{\bf F}}(q)$, and the posterior distribution can be
expressed using Bayes theorem as:\cite{Skilling89,Gubernatis96}
\begin{equation}
P({\bf S}(q)|\tilde{{\bf F}}(q))=P(\tilde{{\bf F}}(q)|{\bf
S}(q))P({\bf S}(q)).
\label{eq:posterior}
\end{equation}
In the above, $P({\bf S}(q))$ is the prior probability distribution
and $P(\tilde{{\bf F}}(q)|{\bf S}(q))$ is the likelihood function.  In
the present study we assume a uniform prior for all positive spectral
functions with a zero moment that obeys the sum rule $\sum
K_{0,j_{\omega}}S_{j_{\omega}}(q)\delta\omega=\tilde{F}_{0}(q)$.
Thus, we can write $P({\bf S}(q))$ as:\cite{Syljuasen08}
\begin{equation}
P({\bf S}(q)) \propto \delta\left(\sum_{j_{\omega}} K_{0,j_{\omega}}
S_{j_{\omega}}(q)\delta \omega - F_{0}(q)\right) \prod_{j_{\omega}}
\Theta (S_{j_{\omega}}(q)),
\label{eq:prior}
\end{equation}
where $\delta(x)$ is Dirac's delta function and $\Theta(x)$ is the
Heaviside step function. The choice of the prior distributions can be
tricky. In order to avoid any undesired bias in the inversion
procedure, we limit the present study to the case of a uniform prior
that meets the necessary sum rule described above.

\begin{figure*}
\begin{centering}
\includegraphics[width=7cm]{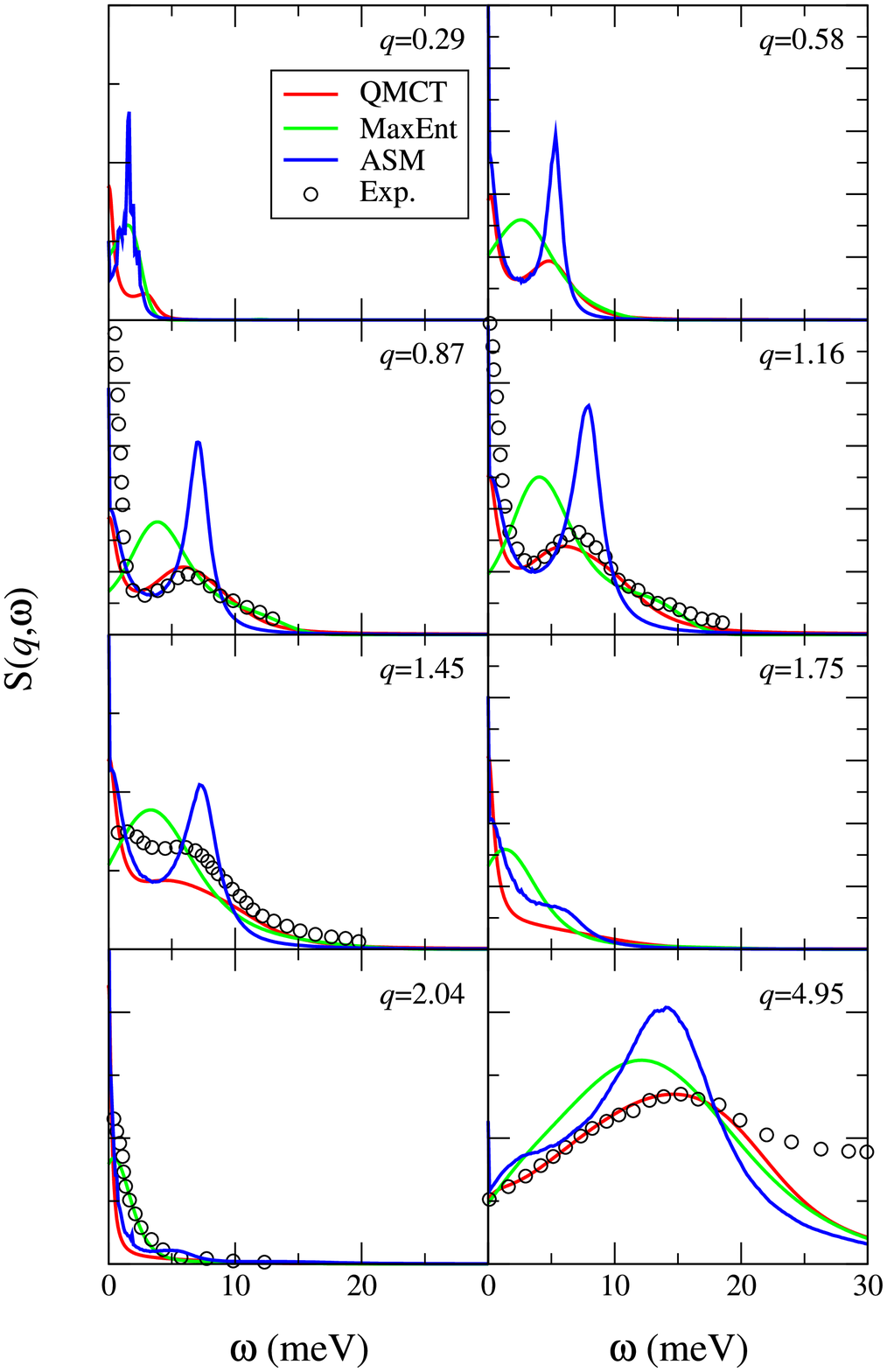}
\includegraphics[width=7cm]{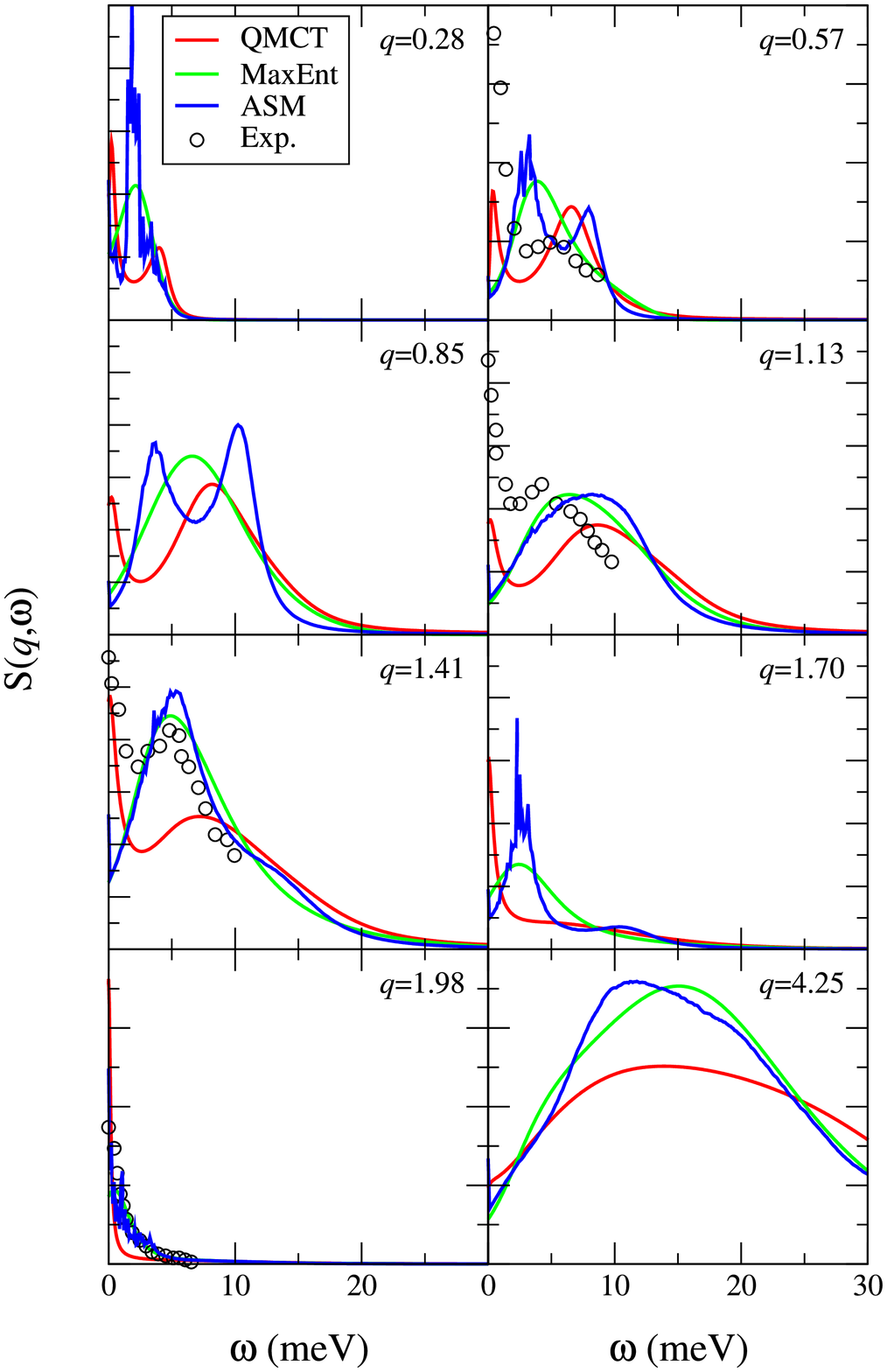} 
\par\end{centering}
\caption{Plots of the dynamics structure factor for liquid {\em
ortho}-deuterium (left panels) and liquid {\em para}-hydrogen (right
panels) for different values of $q$. Experimental results are taken
from Ref.~\onlinecite{Bermejo93} for liquid {\em ortho}-deuterium and
from Ref.~\onlinecite{Bermejo99} for {\em para}-hydrogen. Values of
$q$ are in inverse $\AA$.}
\label{fig:sqw} 
\end{figure*}

The likelihood function appearing in Eq.~(\ref{eq:posterior})
describing the fluctuations of the imaginary time data is assumed to
be of a Gaussian form:
\begin{widetext}
\begin{equation} P(\tilde{{\bf
F}}(q)|{\bf S}(q))\propto\exp\left\{ -\frac{1}{2}Tr
    \sum_{i=1}^{n}(\tilde{{\bf F}}^{i}(q)-\tilde{{\bf F}}(q)^{{\bf
    S}(q)})^{T}{\bf \Sigma}(q)^{-1}(\tilde{{\bf F}}^{i}(q)-\tilde{{\bf
    F}}(q)^{{\bf S}(q)})\right\},
\label{eq:likelihood}
\end{equation}
\end{widetext}
where we have partitioned the PIMC data into $n$ bins each contains
$m$ measurements, the vector $\tilde{{\bf F}}^{i}(q)$ is the average
result for bin $i$, and ${\bf F}(q)^{{\bf S}(q)}$ is the data vector
corresponding to a specific solution ${\bf S}(q)$.  The width of the
Gaussian distribution is taken from the covariance matrix ${\bf
\Sigma}(q)$ with elements of ${\bf \Sigma}(q)$ given by:
\begin{equation}
{\bf \Sigma}_{j_{\tau}k_{\tau}}(q) = \frac{1}{n-1}\sum_{i}
(\tilde{F}_{k_{\tau}}^{i}(q) - \tilde{F}_{k_{\tau}}(q))
(\tilde{F}_{j_{\tau}}^{i}(q) - \tilde{F}_{j_{\tau}}(q)).
\label{eq:sigma}
\end{equation}
The above expression for the likelihood function can further be
simplified~\cite{Syljuasen08} to derive a working expression give by:
\begin{equation}
P(\tilde{{\bf F}}(q)|{\bf S}(q))\propto\exp\left\{ -\frac{n}{2}E({\bf
  S}(q)))\right\},
\label{eq:likelihood-f}
\end{equation}
with the ``energy'' function $E({\bf S}(q))$ given by:
\begin{equation}
E({\bf S}(q)) = Tr(\tilde{{\bf F}}(q)-\tilde{{\bf F}}(q)^{{\bf
S}(q)})^{T}{\bf \Sigma}(q)^{-1}(\tilde{{\bf F}}(q)-\tilde{{\bf
F}}(q)^{{\bf S}(q)}).
\label{eq:energy}
\end{equation}

For readers who are interested in a more comprehensive discussion of
the ASM we refer them to Refs.~\onlinecite{Sandvik98} and
\onlinecite{Syljuasen08}.  Here, we have provided a short overview for
completeness and note that the averaging procedure described above
ensures that solutions (${\bf S}(q)$) which over-fit the noise in the
data ($\tilde{{\bf F}}(q)$) are averaged out, and thus the outcome is
likely to be a smooth spectrum, with realistic features.

\section{Model and results}
\label{sec:res} The major goal of the present study is to conduct
a direct comparison between the ASM, the MaxEnt method, and the QMCT
for collective density fluctuations in liquid {\em ortho}-deuterium
and liquid {\em para}-hydrogen. Both liquids have been studied
extensively over the past decade by many different
techniques.\cite{Rabani02a,Reichman01a,Rabani02b,Reichman02a,Rabani02d,Bermejo00,Zoppi02,Rabani02c,Voth96,Kinugawa98,Zoppi02a,Rabani04,Rossky2004,Voth2004,Rabanireview05,Voth2005,Voth2006,Manolopoulos07,Miller08,Miller08a}
These and other studies show that they are characterized by quantum
dynamical susceptibilities which are not reproducible using classical
theories. Thus, these liquids are ideal to assess the accuracy of
methods developed for quantum liquids and serve as canonical models in
the field.

The static input required by the QMCT and the imaginary time
intermediate scattering function required for the analytic
continuation approaches were generated by PIMC simulations at
$T=20.7K$ and $\rho=0.0254\mbox{\AA}^{-3}$ for liquid {\em
ortho}-deuterium~\cite{Zoppi95} and $T=14K$ and
$\rho=0.0235\mbox{\AA}^{-3}$ for liquid {\em
para}-hydrogen.\cite{Scharf93} The PIMC simulations were done using
the NVT ensemble with $256$ particles interacting via the
Silvera-Goldman potential,\cite{Silvera78,Silvera80} where the entire
molecule is described as a spherical particle, so the potential
depends only on the radial distance between particles.  The staging
algorithm~\cite{Pollock84} for Monte Carlo chain moves was employed to
compute the numerically exact input. The imaginary time interval was
discretized into $N_{\tau}$ Trotter slices of size
$\epsilon=\beta/N_{\tau}$ with $N_{\tau}=20$ and $N_{\tau}=50$ for
liquid {\em ortho}-deuterium and liquid {\em para}-hydrogen,
respectively. Approximately $1-2\times10^{6}$ Monte Carlo passes were
made, each pass consisted of attempting moves in all atoms and all the
beads that were staged. Data collection was taken every $10$ Monte
Carlo steps, block averaging was done for $m=10$ and the total number
of bins was set to $n=1-2\times10^{4}$. The acceptance ratio was
approximately $0.25-0.3$ for both liquids.

In Fig.~\ref{fig:sqw} we show the results for the dynamic structure
factor $S(q,\omega)$ for both liquids and for different values of
$q$. The results are compared to the experiments of Mukherjee {\em et
al.}~\cite{Bermejo93,Bermejo97} for {\em ortho}-deuterium and to the
experiments of Bermejo {\em et al.}\cite{Bermejo99,Bermejo00} for
liquid {\em para}-hydrogen for selected values of $q$ (the theoretical
values of $q$ are slightly different from the experiments due to the
limitations associated with the constant volume simulation and finite
size effects). Several important features are observed experimentally:
First, the present of a finite frequency peak that disappears around
$q=1.4\AA^{-1}$ signifying the presence of collective density
fluctuations (this feature is not observed classically for these
systems). Second, the presence of a low frequency peak associated with
the long-time relaxation of density fluctuations. Finally, as $q$
approaches $q_{max}$ ($q_{max}\approx2\AA^{-1}$ is the value of $q$
where the static structure factor reaches its first maximum) one
observes a quantum mechanical de Gennes narrowing of the dynamic
structure factor.

These features are semi-quantitatively reproduced by the QMCT as shown
in Fig.~\ref{fig:sqw} (a full description of the method can be found
in our previous publication~\cite{Rabani04}). In particular, the
theory captures the position of both the low and high intensity peaks
for liquid {\em ortho}-deuterium and slightly overestimates the
position of the high frequency peak for liquid {\em para}-hydrogen.
The width of the peaks, which are associated with the lifetime of
density relaxation, are also captured by QMCT as is the quantum
mechanical de Gennes narrowing of the dynamic structure factor. The
overall good agreement between the QMCT and the experiments is
remarkable and signifies the importance of the quantum mode-coupling
portion to the memory kernel at long
times.\cite{Rabani02a,Rabani02b,Rabani02d,Rabani04}

Turning to the MaxEnt results for $S(q,\omega)$ (a full description of
the calculation can be found elsewhere~\onlinecite{Rabani04}), it
becomes obvious that MaxEnt fails to provide a quantitative
description of the density fluctuations in these liquids. It is well
known that the MaxEnt approach often fails when several distinct
spectral features appear that are closely spaced in frequency. This is
clearly the case here where the MaxEnt approach predicts a single
frequency peak instead of two, at a position that is approximately the
averaged position of the two experimental peaks. Only when the
dynamics are characterized by a single relaxation time, like the case
at $q_{\max}$, does the MaxEnt approach provides quantitative results.

The failure of the MaxEnt result poses a challenge for analytic
continuation based methods. Can the inversion of Eq.~(\ref{eq:fqtau1})
by other analytic continuation methods produce features that are not
reproducible by MaxEnt? In order to address this problem, we need to
average the spectrum over the posterior distribution, e.g, solve for
the average spectrum as given by Eq.~(\ref{eq:sqwav}). The averaging
procedure is best done by a Metropolis Monte Carlo approach described
in Ref.~\onlinecite{Syljuasen08}.  The basic idea behind the Monte
Carlo procedure is to start with any guess for the spectrum ${\bf
S}(q)$ and then preform a Metropolis walk in the solution space to
obtain different spectra ${\bf S}(q)$ that are distributed according
to posterior distribution $P({\bf S}(q)|\tilde{{\bf F}}(q))$.

\begin{figure*}
\begin{centering}
\includegraphics[width=7cm]{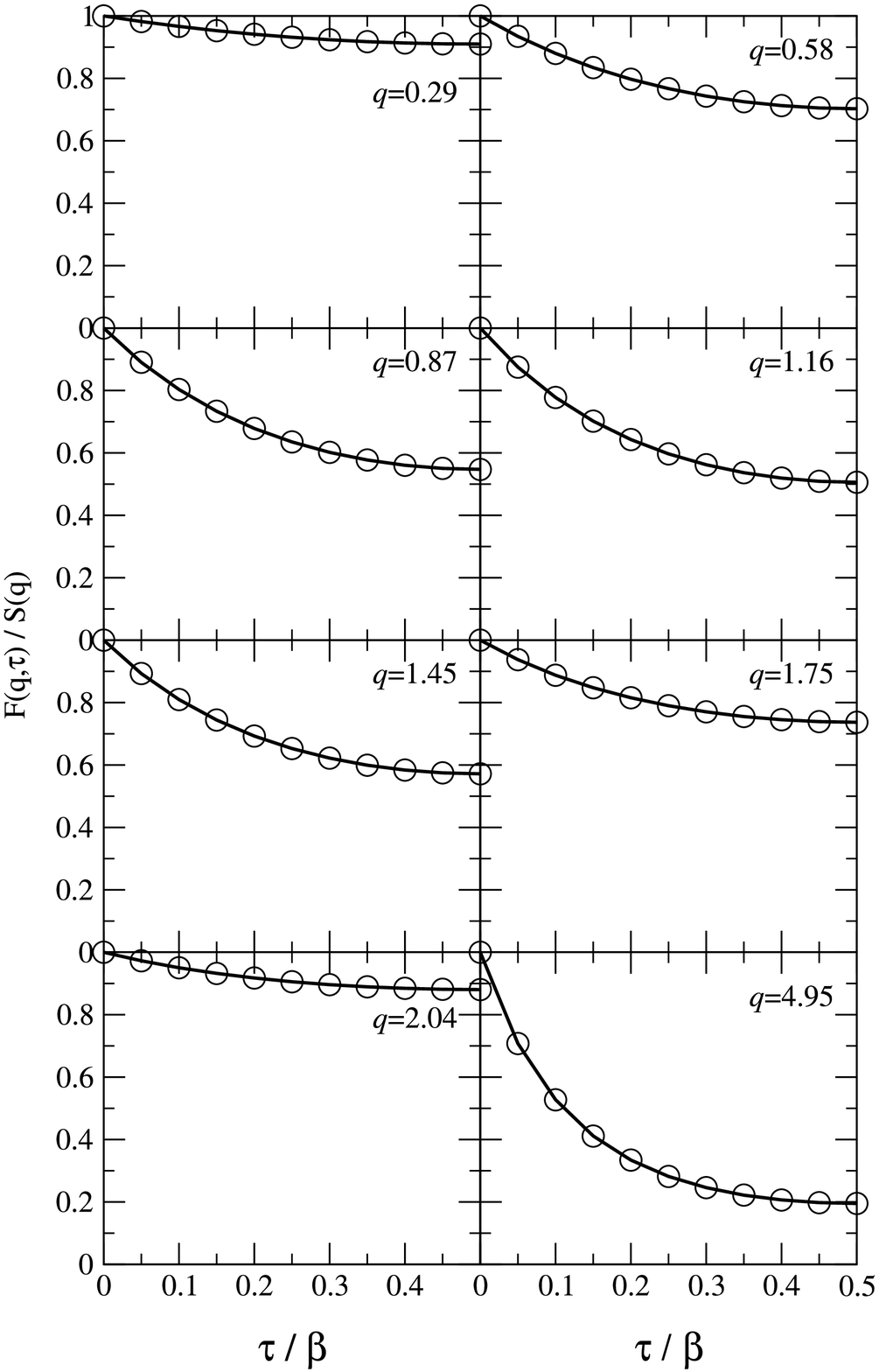}
\includegraphics[width=7cm]{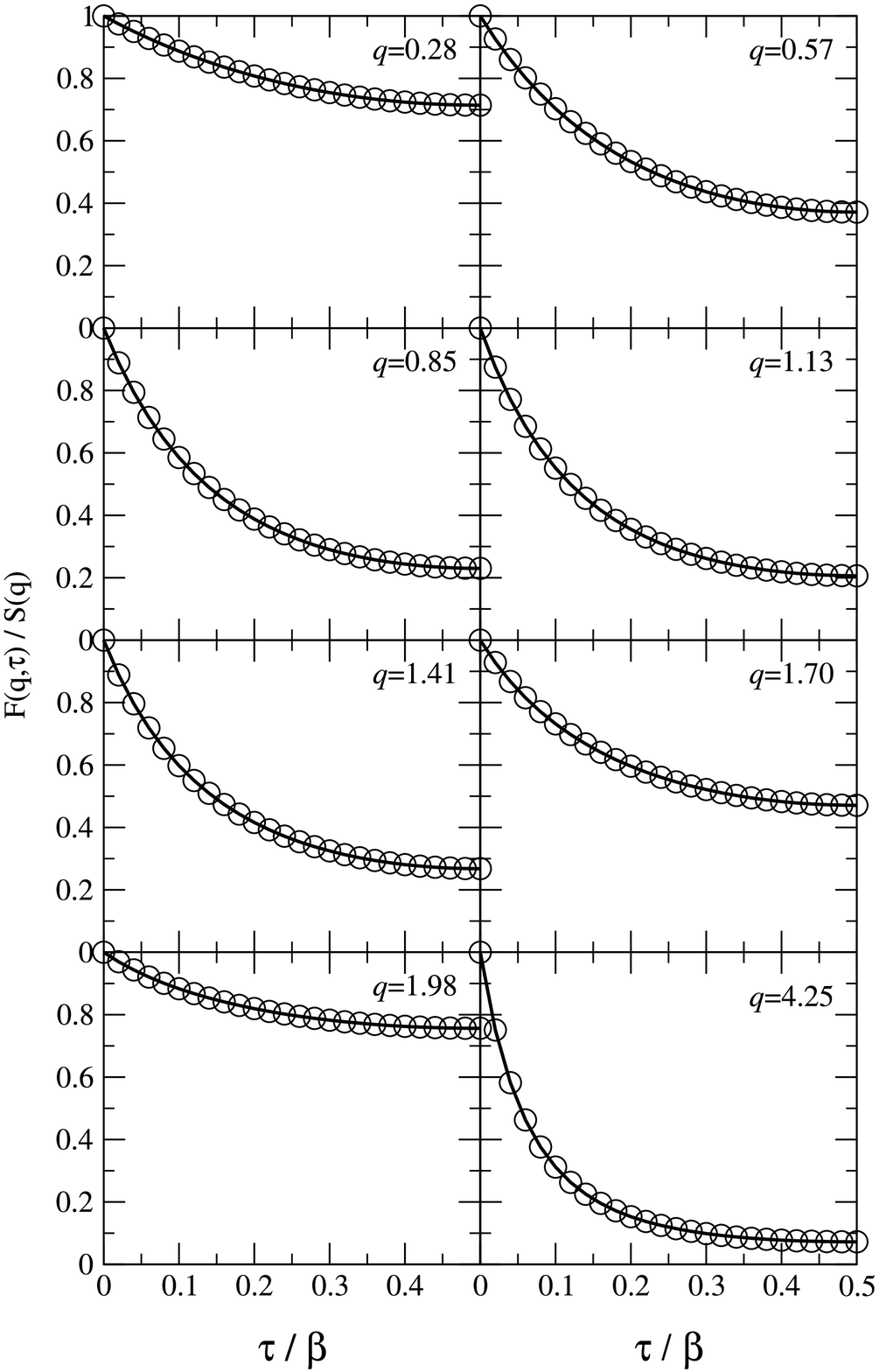} 
\par\end{centering}
\caption{Plots of the imaginary time intermediate scattering function
for liquid {\em ortho}-deuterium (left panels) and liquid {\em
para}-hydrogen (right panels) for different values of $q$. Open
circles represent the results obtained from the PIMC simulations and
the solids lines are fits of the ASM to the PIMC results. Values of
$q$ are in inverse $\AA$.}
\label{fig:fqtau} 
\end{figure*}

In practice, we follow the recipe of Ref.~\onlinecite{Syljuasen08}.
First, we randomly choose a pair of neighboring frequencies,
$j_{\omega}$ and $j_{\omega}+1$. We then select a random number
$\zeta=[-\alpha,\alpha]$ where
$\alpha=S_{j_{\omega}}(q)K_{0,j_{\omega}}+S_{j_{\omega}+1}(q)K_{0,j_{\omega}+1}$,
and make a trial move that preserve the sum rule $\sum
K_{0,j_{\omega}}S_{j_{\omega}}(q)\delta\omega=\tilde{F}_{0}(q)$:
\begin{equation}
\begin{split}
S_{j_{\omega}}^{new}(q)=S_{j_{\omega}}(q)+\zeta\frac{K_{0,j_{\omega}+1}}{K_{0,j_{\omega}}+K_{0,j_{\omega}+1}}\\
  S_{j_{\omega}+1}^{new}(q)=S_{j_{\omega}+1}(q)-\zeta\frac{K_{0,j_{\omega}}}{K_{0,j_{\omega}}+K_{0,j_{\omega}+1}}
\end{split}
\label{eq:move}
\end{equation}
The trial move is accepted only if $S_{j_{\omega}+1}^{new}(q)\ge0$
with the Metropolis probability given by:
\begin{widetext}
\begin{equation}
P({\bf S}(q)\rightarrow{\bf S}^{new}(q))=P({\bf
S}^{new}(q))\mbox{min}\left\{ 1,\exp\left[-\frac{n}{2}\left(E({\bf
S}^{new}(q))-E({\bf S}(q))\right)\right]\right\},
\label{eq:accept}
\end{equation}
\end{widetext}
where $E({\bf S}(q))$ is given by Eq.~(\ref{eq:energy}).  In the
present study we employ a simple Metropolis algorithm to sample the
values of ${\bf S}(q)$ . We did not encounter difficulties with
sampling that require more sophisticated Monte Carlo procedures like
parallel-tempering.\cite{Syljuasen08} As a check, we started the Monte
Carlo procedure from $10$ different initial conditions. All the
different runs converged to the same solution within the statistical
noise of the Monte Carlo sampling. The ASM results shown in
Fig.~\ref{fig:sqw} where averaged over $\approx10^{8}$ Monte Carlo
sweeps, each consist of an attempt to change $N_{\omega}=512$ solution
points.

The main and perhaps, unexpected, result is that the ASM captures
qualitatively all feature observed experimentally for the dynamics
structure factor, as shown in Fig.~\ref{fig:sqw}. Contrary to MaxEnt,
the ASM reproduces the low and finite frequency peaks, albeit the fact
that it somewhat underestimates the width of the finite frequency peak
and also slightly overestimates the position of the low frequency
peak. Overall, the ASM agrees better with the QMCT and with the
experimental results as compared to MaxEnt. The fact that there are
features that are observed experimentally and captured by QMCT and ASM
and not by MaxEnt method, indicate the superiority of the former
approaches over MaxEnt.

Despite the semi-quantitative agreement of the QMCT and ASM approaches
with experiments, there are still significant differences, as clearly
is the case for liquid {\em para}-hydrogen. It still remains a
challenge to determine the accuracy of both approaches since an exact
solution for this many-body quantum mechanical problem is still (and
may always be) out of reach. A direct comparison with experiments can
be misleading, since several approximations are made for both QMCT and
ASM. First, both are based on the assumption that the dynamics evolve
over the potential energy taken from the work of Silvera and
Goldman.\cite{Silvera78,Silvera80} Furthermore, for simplicity, {\em
para}-hydrogen and {\em ortho}-deuterium were treated as spherical
particles, while experimentally this is not the case. Finally, for
convenience, the simulations were done for the NVT ensemble while
experiments are conducted at constant pressure (NPT ensemble). These
approximation may lead to the difference observed between the theory
and experiments, but the differences may also be attributed to the
approximations made by QMCT and ASM for the dynamics
itself. Nonetheless, the encouraging agreement between QMCT and ASM
provides stronger support for their accuracy.

Before we conclude, in Fig.~\ref{fig:fqtau} we show the excellent
agreement between the imaginary data generated by the PIMC simulations
and imaginary data obtained by the inversion of Eq.~(\ref{eq:fqtau1}).
As is well understood, only disagreement at this level can be used to
draw decisive conclusion about the inversion process. However, it is
quite satisfactory that the Monte Carlo procedure in the ASM leads to
an excellent agreement for the imaginary time data.

\section{Conclusions}
\label{sec:conclusions}
A comparison between the QMCT, MaxEnt method, and ASM has been made
for the case of collective density fluctuations in liquid {\em
ortho}-deuterium and liquid {\em para}-hydrogen.  As far as we know,
the present study is the first attempt to apply the ASM to dynamical
susceptibilities for off-lattice models. We find that the main
features observed experimentally for the cases discussed above are
captured by QMCT and ASM, but not by MaxEnt. The results obtained by
the ASM appear to coincide with the experimental results more closely
in the case of {\em ortho}-deuterium. In this case the ASM produces
results that are narrower than those seen in experiment. However, it
should be expected that the theoretical spectrum is indeed sharper
than that obtained in experiment because no instrumental broadening
function is included in the theoretical calculations. The better
agreement in the case of {\em ortho}-deuterium compared to {\em
para}-hydrogen for the ASM may be a result of the better statistical
quality of the data used in the {\em ortho}-deuterium case.

Our ASM results for the non-trivial examples presented here give
strong motivation to continue the investigation of the ASM approach in
other systems. One important problem worthy of reinvestigation is
study of density fluctuations in superfluid helium. Boninsegni and
Ceperley~\cite{Ceperley96} studied this problem and found that the
sharp features associated with roton peaks could not be reproduced by
the MaxEnt approach. It would be interesting to see what the ASM
yields for this challenging and important problem.  Indeed, since the
the ASM results presented here and by others provide clear evidence
that analytic continuation based approaches are not limited to the
case where dynamical susceptibilities are characterized by a single
timescale, and since the ASM is a flexible approach suitable to the
study of any type of correlation function for which imaginary time
quantum Monte Carlo data can be generated, we expect that this
approach will be further developed and utilized as a powerful means of
extracting dynamical information from imaginary time simulations.

We would also like to add a word that at this stage it is still not
entirely clear what should be trusted {\em a priori} in the ASM
results and if the ASM is universally superior to MaxEnt.  Unpublished
work investigating the behavior of both MaxEnt and the ASM in lattice
models of the temperature and doping dependence near the Mott
transition actually suggest that MaxEnt can in some cases provide a
superior description of certain spectral features, but is inferior in
its description of others.\cite{Me} For example MaxEnt fails
completely in describing the Fermi liquid behavior of the self energy
in the low frequency range while, rather surprisingly, the ASM
does. Thus, we may conservatively suggest that at the very least, the
ASM provides a complimentary and useful tool for extracting real time
features from imaginary time quantum Monte Carlo data that appears to
be superior to MaxEnt at least in its ability to uncover sharp
spectral features. Assessing the accuracy of the ASM in a more
systematic manner is another worthy direction of future research.

\section{Acknowledgements}
DRR would like to thank the NSF for funding and Andy Millis for useful
discussions.

\end{document}